\documentclass[trackchanges]{aastex701}
\usepackage[utf8]{inputenc}  
\usepackage{amsmath}         
\usepackage{amsfonts}        
\usepackage{amssymb}         
\usepackage{cleveref}

\begin{document}

\title{The Sesquinary Catastrophe on Deimos can reconcile its excited past with its dynamically cool present.}

\correspondingauthor{Kaustub P. Anand}
\email{anand43@purdue.edu}

\author[0009-0007-2467-0139]{Kaustub P. Anand}
\affiliation{Department of Physics and Astronomy, \\
Purdue University, \\
525 Northwestern Ave, West Lafayette, IN 47907, USA}
\affiliation{Department of Earth, Atmospheric, and Planetary Sciences, \\
Purdue University, \\
550 Stadium Mall Drive, West Lafayette, IN 47907, USA}
\email{anand43@purdue.edu}

\author[0000-0003-1226-7960]{Matija \'Cuk}
\affiliation{SETI Institute, 339 N Bernardo Ave, Suite 200, Mountain View, CA 94043, USA} 
\email{mcuk@seti.org}

\author[0000-0003-1656-9704]{David A. Minton}
\affiliation{Department of Earth, Atmospheric, and Planetary Sciences, \\
Purdue University, \\
550 Stadium Mall Drive, West Lafayette, IN 47907, USA}
\affiliation{Department of Physics and Astronomy, \\
Purdue University, \\
525 Northwestern Ave, West Lafayette, IN 47907, USA}
\email{daminton@purdue.edu}

\begin{abstract}

The origins of the Martian moons Phobos and Deimos are highly debated, and hypotheses include formation from an impact-generated circum-Martian disk or from capture of asteroids. With the impact scenario, Deimos (or its precursors) were formed or were pushed out beyond the synchronous orbit of Mars. Moons interior to the synchronous orbit, including Phobos (or its precursors), would tidally evolve and resonances between these moons could potentially excite Deimos' orbit. This contradicts Deimos' present-day orbit of low eccentricity ($0.00027$) and moderate inclination ($1.8^\circ$ to the Laplace plane). Tidal dissipation within Deimos is too inefficient for eccentricity damping, and without alternative mechanisms, Deimos' present-day orbit places strong constraints on the evolution of any inner moons. We propose that a runaway collisional cascade called the ``sesquinary catastrophe'' acts as a natural barrier that prevents Deimos from having a more excited orbit. Using N-body simulations with collisional fragmentation, we show that if Deimos was more excited, it would undergo a sesquinary catastrophe and break apart into a Roche-exterior debris disk. Using a measure of sesquinary orbital excitation called $q$, our simulations and previous works suggest that breakup occurs for $q \gtrsim 8$ on timescales of $\sim 10^{3-4}$~years. If Deimos was destroyed in a sesquinary catastrophe and re-accreted from a (likely collisionally) damped debris disk, it should be a porous sand-pile moon, consistent with its smooth surface. The sesquinary catastrophe can be applied to other Deimos-like planetary moons at $q \gtrsim 8$.

\end{abstract}



\section{Introduction}  \label{sec:intro}
The origin of the Martian moons Phobos and Deimos is intensely debated with hypotheses ranging from an intact capture of two asteroids~\citep{Hunten1979CaptureDrag, Tolson1978VikingResults, Pang1978TheAnalysis, Pollack1978MulticolorComposition, Rivkin2002Near-InfraredDeimos} to capture and breakup of a single body~\citep{Kegerreis2024OriginAsteroid, Bagheri2021DynamicalProgenitor} to a giant impact that produces a circum-Martian disk~\citep{Burns1992ContradictoryMoons., Hyodo2022ChallengesMoon, Hesselbrock2017AnDeimos, Canup2018OriginMars, Rosenblatt2016AccretionMoons, Rosenblatt2012OnDisk, Craddock2011AreImpact, Citron2015FormationImpact, Hyodo2017OnAspects, Hyodo2017OnEvolution}. The near-equatorial orbits of Phobos and Deimos (see \cref{tab:moons orbital el}) would naturally result
from formation from a flat disk. Because of this, the theory of a giant impact is arguably the most widely accepted at this time, although the possibility that the disk was made of captured material~\citep{Kegerreis2024OriginAsteroid} cannot be excluded. However, as a consequence of the dynamical evolution of a multi-moon system created in a giant impact~\citep{Rosenblatt2012OnDisk}, proto-Deimos is likely to be left on a dynamically excited orbit. This excitation contradicts the extremely low eccentricity of modern-day Deimos ($e = 0.00027$~\citep{Jacobson2014MartianEphemerides}), leaving the giant impact formation theory of Phobos and Deimos in doubt. Note that the inclination of Deimos may have originated in a much later resonance with an inner moon~\citep{Cuk2020EvidenceDeimos}, which was a part of a long-term ring-moon cycle at Mars~\citep{Hesselbrock2017AnDeimos}, but such a dynamical event would postdate the interaction between the first-generation moons generated in the giant impact by hundreds of Myr, or even Gyrs. 

In this work, we show that the sesquinary catastrophe~\citep{Cuk2023SesquinaryOrbits} can reconcile an excited past Deimos with its dynamically cool present. A runaway cascade of sesquinary impactors~\citep{Zahnle2008SecondaryEuropa} can break apart an excited moon, lead to a ring of smaller debris that can be more easily circularized, and then re-accrete into a dynamically cooler moon~\citep{Anand2024SesquinaryDeimos, Anand2025ThePresent.}. 
This paper is set up as follows. Section \ref{sec:background} discusses relevant background information. The methods of our work are described in section \ref{sec:methods} with the results in section \ref{sec:results}. The implications of our results are discussed in section \ref{sec:discussion} with conclusions in section \ref{sec:conclusions}.

\section{Background} \label{sec:background}

Phobos and Deimos are both small ($R<20$~km), irregular in shape, and relatively close-in to their host planet Mars ($a<7 R_M$). Phobos, the inner moon, is inside the synchronous orbit ($R_{sync} \simeq 6 R_M$) and is tidally evolving inwards rapidly towards Mars~\citep{Black2015TheSystem}. On the other hand, Deimos is just outside the synchronous orbit and is very slowly evolving away from Mars. The formation history of Phobos and Deimos is puzzling for many reasons. Their reflectance spectra is largely featureless and flat with a low albedo, matching that of dark, carbonaceous asteroids, which has lead to the hypothesis that that they are captured asteroids~\citep{Rivkin2002Near-InfraredDeimos,Tolson1978VikingResults, Pollack1978MulticolorComposition}. However, flat featureless spectra are not unique to carbonaceous asteroids, and recent work has shown that Phobos may have a basaltic component, which implies some mixing with Mars' crust~\citep{Glotch2018MGS-TESPhobos}. In addition, space weathering can flatten basaltic reflectance spectra on small airless bodies~\citep{Pieters2016SpaceBodies}. Phobos is also being bombarded by oxygen ions from the Martian atmosphere~\citep{Dong2015StrongChannel, Nenon2019PhobosObservations} and oxygen ion irradiation experiments on on Phobos simulant material shows altered spectral features~\citep{Tabata2025EFFECTIONS}. 

However, the moons' low eccentricity and near-equatorial orbits (see \cref{tab:moons orbital el}) contradict a captured origin and instead imply an in-situ origin from a circum-Martian disk~\citep[][and others]{Burns1992ContradictoryMoons., Rosenblatt2016AccretionMoons, Hesselbrock2017AnDeimos, Canup2018OriginMars}. While most works favor a debris disk from a giant impact, \citet{Kegerreis2024OriginAsteroid} recently suggested that the disk could be made up of an asteroid broken-up by Mars' tidal field. Lastly, \citet{Bagheri2021DynamicalProgenitor} proposed that Phobos and Deimos were formed as collisional fragments of a single moon. The fragments would initially have highly eccentric and near-equatorial orbits, but this hypothesis requires implausible ejection velocity fields, highly divergent tidal properties for the two moons, and the pair avoiding immediate re-impact while still on crossing orbits~\citep{Hyodo2022ChallengesMoon}. Later in the paper we will address the likely impossibility of Phobos and Deimos maintaining very eccentric orbits without triggering a sesquinary catastrophe.

\begin{deluxetable}{cccccccc}
\tablenum{1}
\tablecaption{Physical and orbital parameters of Phobos and Deimos.\label{tab:moons orbital el}}
\tablehead{\colhead{Moon} & \colhead{Radius (km)} & \colhead{Mass ($10^{15}$~kg)} & \colhead{Semi-major axis ($R_M$)} & \colhead{Eccentricity} & \colhead{Inclination ($^{\circ}$)} & \colhead{$q$} & \colhead{$\tau$ (years)}}
\startdata 
Phobos & 11 & 10.8  &  2.762 & 0.0151 & 1.08 & 4.6 & 38\\
Deimos & \phantom{0}6 & \phantom{0}1.8 & 6.911 & 0.0003 & 1.79 & 7.6 & 1600\\
\enddata
\tablecomments{$\tau$ is the sesquinary re-impact timescale defined by~\citet{Cuk2023SesquinaryOrbits} and in \cref{eqn:tau}}
\tablerefs{\citet{Murray2000SolarDynamics} for the radii and masses; \citet{Brozovic2025RevisedDeimos} for the mean orbital elements where Mars radius $R_M=3394$~km; \citet{Cuk2023SesquinaryOrbits} for sesquinary excitation $q$ and collisional timescale $\tau$.}
\end{deluxetable}

Multiple large moons can form out of the disk after the giant impact~\citep{Rosenblatt2016AccretionMoons, Canup2018OriginMars}. These inner moons (interior to the synchronous) tidally evolve inwards towards Mars and progressively fall onto it with Phobos possibly being the last surviving inner moon. Alternatively, the circum-Martian disk can subsequently undergo multiple ring-moon cycles~\citep{Hesselbrock2017AnDeimos}, and Phobos may be the last representative of this process. In both models, the outward migration of the inner moons due to disk torques from the impact-generated disk and the inward migration of the inner moons due to tides is considered. The latter is more prominent when the original impact-generated disk has dissipated enough. In the ring-moon cycling model, the inner moons first evolve outwards away from Mars due to Lindblad torques with the ring it formed from. Then they evolve inwards towards Mars once the ring has lost enough mass that the outward Lindblad torques are no longer strong enough to counteract the inward tidal torque. The inner moons get close to Mars and break apart, forming a ring, and restarting this cycle. Regardless of the exact sequence of events, these models are able to recreate modern-day Phobos, but not Deimos.


\subsection{Solving Deimos' excited past with the sesquinary catastrophe} \label{sec:sesquinary intro}

As the first generation of moons are formed from an impact-generated disk, Deimos has to be placed beyond the synchronous orbit by a mechanism other than tidal migration. Deimos could either be a remnant of the original impact~\citep{Hesselbrock2017AnDeimos, Canup2018OriginMars} or was pushed out by mean-motion resonances with the inner moons~\citep{Rosenblatt2016AccretionMoons}. As the large inner moons are formed and then initially migrate outwards due to interaction with the disk, they enter into orbital resonances with Deimos that tend to excite Deimos's eccentricity or inclination~\citep{Malhotra1993TheOrbit} beyond its current values. In addition, \citet{Canup2018OriginMars} argue that Deimos is even destabilized and removed from the system if the initial debris disk is too massive ($M_{Disk} \ge 10^{22}$~g). The small mass of Deimos also makes it somewhat difficult to simulate it directly from giant impact models due to resolution limitations. \citet{Cuk2020EvidenceDeimos} also show that an outward moving large Phobos precursors can excite Deimos' inclination upto $3.5^\circ$. As a result, previous works either cannot form modern-day Deimos, or if they do, they form a dynamically excited Deimos that does not match its current orbit~\citep{Rosenblatt2016AccretionMoons, Hesselbrock2017AnDeimos, Canup2018OriginMars, Cuk2020EvidenceDeimos}\footnote{While \citet{Canup2018OriginMars} do not give eccentricity estimates for Deimos, our simulations to recreate their results show Deimos analogs excited up to $e \sim 0.03$. \citet{Rosenblatt2016AccretionMoons, Hesselbrock2017AnDeimos} mention the excitation issues with recreating Deimos.}. However, a runaway erosion of sesquinary impactors, called the sesquinary catastrophe, can help de-excite Deimos to its current orbit, and therefore the present-day level of orbital excitation is not as strong of a constraint on its orbital history~\citep{Cuk2023SesquinaryOrbits}. 

This runaway erosion process can break up small, close-in, and excited moons~\citep{Cuk2023SesquinaryOrbits}. Sesquinary impactors are ejecta from a large (likely heliocentric) impact on a satellite that escape its gravitational pull, orbit the host planet on an independent orbit, and then re-impact the original satellite~\citep{Zahnle2008SecondaryEuropa}. From \citet{Cuk2023SesquinaryOrbits}, sesquinary impactors impact the original moon at $v_{impact} \sim q \times v_{escape}$ where $q$ is sesquinary excitation: 

\begin{equation} \label{eqn:q}
    q = \sqrt{e^2 + sin^2(i)}\frac{v_{orbital}}{v_{escape}}
\end{equation}  

where $v_{orbital}$ is the orbital velocity, $v_{escape}$ is the escape velocity of the moon and $e$ and $i$ are eccentricity and inclination of the satellite. As the ejecta escape the satellite, they take on effectively the same orbit as the satellite. The typical re-impact time for sesquinary impactors is

\begin{equation} \label{eqn:tau}
    \tau = \frac{P}{\Delta}\frac{a^2 \delta e \delta i}{R^2}
\end{equation} 

where $P$~is orbital period of the moon, $\Delta = \frac{v_{escape}}{v_{orbital}}$, $R$~is the radius of the moon, $a$ is the semi-major axis, and $a^2 \delta e \delta i$ is the ``box" that the impactors can stay in~\citep{Cuk2023SesquinaryOrbits}. If a given satellite is small (low $v_{escape}$), close-in to the planet (high $v_{orbital}$), and has noticeable excitation (high $e$ and/or $i$), sesquinary impacts can happen at a high velocity once the orbits of the moon and ejecta have precessed out of alignment. These impacts can then launch more ejecta off the surface of the satellite, which in turn become projectiles that re-impact the original satellite, release more debris, creating a positive feedback loop that grinds down the excited moon. Note that as the moon loses mass, its $v_{escape}$ will decrease, and increase its excitation $q$ over time. Currently, Deimos has a $v_{escape} \sim 5.56$~m/s, $q \sim 7.6$, $\tau \sim 1600$~years~\citep[see \cref{tab:moons orbital el}, ][]{Cuk2023SesquinaryOrbits}.

\section{Methods} \label{sec:methods}

To test the viability of the sesquinary catastrophe, we ran N-body simulations using Swiftest~\citep{Wishard2023Swiftest:Systems} with the Symplectic Massive Body Algorithm (SyMBA) integrator~\citep{Duncan1998AENCOUNTERS} and the in-built gravitational harmonics capability. SyMBA can correctly handle close encounters, and thus collisions, between massive particles (impactor-impactor and moon-impactor). We set up a Mars-centric system with Phobos, Deimos, and the sesquinary impactors. All bodies are set up as massive particles that can collide and fragment.

\subsection{Collisional model} \label{sec:collisional model}

Swiftest has an in-built collisional fragmentation model called FRAGGLE that conserves angular momentum and follows a final size distribution derived from \citet{Leinhardt2012COLLISIONSLAWS}~\citep{Wishard2023Swiftest:Systems}. FRAGGLE follows scaling laws from \citet{Leinhardt2012COLLISIONSLAWS} for collisions between similar-sized bodies ($M_{impactor} / M_{target} > 1/500$) and from \citet{Hyodo2020EscapeEjecta} for cratering collisions ($M_{impactor} / M_{target} \leq 1/500$) for the properties of the resultant impact ejecta. The code decides the appropriate collisional regime based on the mass ratio of the impactor and target involved in a given impact. Because sesquinary impactors are much smaller than the target moon, the cratering model is more appropriate and ends up being used for the bulk of our work. 

We incorporate two different models into Swiftest because each set of scaling laws is defined only for a given collisional regime. Using the scaling from one regime in another gives unrealistic values. For example, when using the scaling laws from \citet{Leinhardt2012COLLISIONSLAWS}, our test simulations showed that an impactor on Deimos with a mass ratio $M_{impactor} / M_{target} = 1/27000$ needs an impact velocity of $\sim 50-100 \ v_{escape}$ to cause any of kind of mass loss from Deimos. This is an extremely high and rather unrealistic velocity for planetocentric impacts. \citet{Cuk2023SesquinaryOrbits} and \citet{Leinhardt2012COLLISIONSLAWS} show that erosion from impacts should occur for $v_{impact} \gtrsim 5-10 v_{escape}$. \citet{Hyodo2020EscapeEjecta} tests and defines the scaling laws for collisions in the cratering or small impactor regime. Using this model, we typically see net mass loss from the target body at $v_{impact} \gtrsim 5 v_{escape}$ (impact angle averaged). This is more in line with what is expected.

The scaling laws in \citet{Hyodo2020EscapeEjecta} are defined for bodies in the gravity regime that are larger and have faster impact velocities than those in the Martian moon system. Without more applicable collisional models publicly available, we find it appropriate to use these laws in this work. We justify this because the scaling laws are normalized to impactor mass and escape velocity units, the impactors and target body in our work also lie in the gravity regime (as defined in \citet{Stewart2009Velocity-DependentPlanetesimals}), and have similar mass ratios tested in \citet{Hyodo2020EscapeEjecta}.  Using these scaling laws, we obtain the final ejecta velocity spread and total mass lost from the target body as a function of impact velocity and impact angle.

\subsection{Initial Setup} \label{sec:sim set up}

As mentioned earlier, we use a Mars-centric system with Phobos, Deimos, and sesquinary impactors. All bodies are set up as massive particles to allow for collisional fragmentation. Because of the size difference and computational constraints, the sesquinary impactors are set up as semi-interacting particles. Semi-interacting particles are massive particles that can gravitationally affect, collide, and fragment with bigger bodies (above a set mass threshold), but do not interact with each other. This allows us to gather the same information about small sesquinary impactors affecting a larger Deimos without adding the extra runtime and data of pairwise impactor interactions. The debris disk is gravitationally dominated by Deimos. As a result, the impactor-impactor interactions are small and do not affect the large scale results. This is corroborated by simulations with the impactors as fully massive particles, i.e., they can collide, fragment, and interact with each other as well. These simulations showed no noticeable difference in results, but were significantly slower by about 2 orders of magnitude and generated $\sim 20-40 \times$ the collisional fragments when compared to simulations with the same set up with semi-interacting impactors. In addition, most of the collisional fragments generated (and subsequent impactors) are smaller and almost all become semi-interacting particles as they fall under the mass threshold within $\sim 500$~y of the simulation anyway.

We set up the Mars-centric system as shown in panel (a) of \Cref{fig:sesquinary sim plots}. We use ephemerides from the JPL Horizons service\footnote{JPL Horizons API: https://ssd.jpl.nasa.gov/horizons/. First accessed in June 2024 with the same epoch date of 1976-08-05 used for all simulations.} for Phobos and Deimos and then add gravitational harmonics (including tesseral terms) from GMM3~\citep{Genova2016SeasonalScience} up to degree $l = 6$ for Mars. Deimos is then given an excitation, i.e., high $e$ and/or $i$, to test the sesquinary catastrophe at various excitations from $0 \leq q \leq 40$. We run about $90$ simulations with arbitrary combinations of $e$ and $i$ for the excited Deimos per simulation bounded by $q \leq 40$. Higher values of $q$ are unlikely and only seen in moons in resonances, except Pallene~\citep{Cuk2023SesquinaryOrbits}. The range of tested eccentricities are $0 \leq e \leq 0.075$ and tested inclinations are $0^\circ \leq i \leq 10^\circ$. 

We do not mean to imply a time for when the sesquinary catastrophe happened to Deimos because we used modern values for Phobos and Deimos. Deimos is very close to the synchronous orbit and its semi-major axis has not changed much due to tides~\citep{Burns1978TheMoons, Cazenave1980OrbitalOrigin}. In addition to this, Deimos' past orbit is chaotic because of planetary perturbations and the Martian obliquity with no major effect from Phobos~\citep{Cuk2025TwoPhobos}. Phobos, on the other hand, is tidally moving inward much quicker on Myr timescales~\citep{Black2015TheSystem} and its location matters more. However, Phobos' history is highly debated with it either being formed at $3.2$~$R_{Mars}$ close to its current location or further out  between $5.2 - 6$~$R_{Mars}$ inside the synchronous orbit~\citep{Cuk2025TwoPhobos} and/or having multiple precursors~\citep{Hesselbrock2017AnDeimos}. As Phobos evolves inwards it crosses mean motion resonances (MMRs) with Deimos. If Deimos and the debris are excited enough, it could potentially trigger another sesquinary catastrophe and this does not change the mechanism we are studying. The resonance timescales are on the order of $10^5$~y, much longer than our simulation timescales of $10^3$~y, and excitations from modern Phobos are not very high for Deimos (up to $e = 0.0007$ from the 3:1 MMR,~\citet{Cuk2025TwoPhobos}) regardless. Phobos also does not affect the scattering or behavior of debris particles around Deimos in our simulations on $10^3$~y timescales. Testing the effect of different locations of Phobos on debris particles is left for future work. As a result, we decided that using modern Phobos and Deimos values are a justifiable choice. 

We then isotropically place and launch $\sim 200$ ejecta particles off of the surface of Deimos at $1 - 3 \ v_{escape}$ with varying individual particle radii. The initial velocity is pointing radially away from Deimos and the initial location of the ejecta on Deimos' surface shows no large-scale difference in results. The position and velocity vectors are converted from a Deimos-centric to the Mars-centric frame and then we allow the system to evolve. The total ejecta mass is typically $10^{-3} \ M_{Deimos}$ (or $0.1\% M_{Deimos}$, but depends on the nature of the simulation (see \Cref{fig:mass loss sims vs q,fig:mass loss sims with mass steps}). The initial total impactor mass is driven by $10^{-4} \ M_{Deimos}$, the expected amount of material expelled from Deimos by a Voltaire-causing impact~\citep{Nayak2016EffectsGeology}. Voltaire is the largest crater on Deimos. We decided to set the initial impactor mass higher than that to account for other impacts that have occurred on Deimos. While this is a slightly arbitrary choice, the dynamics remain the same albeit with a longer timescale (see section~\ref{sec:breakup timescale results}). The orbits of the ejecta precess out of alignment and eventually re-impact Deimos as sesquinary impactors. These simulations had a timestep of $0.003$~days and were run for $5000$~years. 

We decided to start with $200$ particles from trial and error. As the system evolves, a large number of collisional fragments are generated and most simulations slow down considerably (by about 2 orders of magnitude in walltime/step with $\sim 10^5$ particles). We are then unable to generate efficient and meaningful results because of the reduced speed and large data size. Typical simulations end up running for $\sim 2000 - 4000$~years. Because we are unable to simulate the system from start to finish, i.e., erode Deimos completely, we use the simulation data to guide a semi-analytical approach.

\subsection{Semi-analytical timescale} \label{sec:semi-analytical timescale}

From the simulation results, we see slow mass loss at various initial $q$ values. We can then semi-analytically estimate a breakup timescale as a function of $q$ from the mass loss rate $dM/dt$. To do this, we will use the re-impact time in \cref{eqn:tau}~\citep{Cuk2023SesquinaryOrbits}, mass scaling laws~\citep{Hyodo2020EscapeEjecta}, and impact velocity distribution from simulations. We start by simplifying the re-impact time for sesquinary impactors $\tau$ (\cref{eqn:tau}) to a function of $M$. Substituting the standard formulas for orbital period $P, v_{escape}, \text{and } v_{orbital}$ results gives us
\begin{equation} \label{eqn:tau simplified}
    \tau = K/M
\end{equation}
with a constant $K$ where

\begin{align} \label{eqn:K (tau constant)}
        K & = \sqrt{\frac{2\pi \rho M_{Planet}}{3a}}\times a^2 \delta e \delta i\ \text{(Orbital Period units)} \\
    & = \sqrt{\frac{(2\pi)^3 \rho}{3G}} \times a^3 \delta e \delta i\ \  \text{ (SI units)}\\
\end{align} 

Here $\rho$ is density of the moon, $G$ is the universal gravitational constant, and $M_{planet}$ is the mass of the host planet. Next, we look at the mass scaling laws from \citet{Hyodo2020EscapeEjecta}. The mass change per impact ($\delta M$) for a given $v_{impact}$ is defined in impactor mass units ($m_{i}$) and depends on $v_{impact} / v_{escape}$ and impact angle $\theta$. Our simulations account for the angle per impact when using these scaling laws. Because our simulations show that the impacts on Deimos in our simulations uniformly span the whole range of $0^\circ - 90^\circ$, we use the angle-averaged mass scaling laws (equations 11 and 12 in \citet{Hyodo2020EscapeEjecta} and below) in the derivation of the semi-analytical timescale:

\begin{equation} \label{eqn:scaling laws form}
\delta M(\eta) = 
\begin{cases}
0.02 \, \eta^{2.2} + 0.071 \, \eta^{0.88} - 0.85, & \text{if } 0 \leq \eta < 12 \\
0.076 \, \eta^{1.65} + 0.071 \, \eta^{0.88} - 0.85, & \text{if } 12 \leq \eta < 16.79 \\
0.076 \, \eta^{1.65}, & \text{if } \eta \geq 16.79
\end{cases}
\end{equation}

where for brevity, we define $\eta = v_{impact}/v_{escape}$. This leaves $\delta M$ as a function of impact velocity only. To average the mass loss from equation \ref{eqn:scaling laws form}, we must obtain a distribution of $\eta = v_{impact} / v_{escape}$ as a function of initial $q$. From \citet{Cuk2023SesquinaryOrbits}, 

\begin{align} 
    v_{impact} & \sim q\times v_{escape} \label{eqn:v_impact f(q)}\\
    \Rightarrow \ \eta & \sim q \label{eqn:eta sim q}
\end{align}
but the exact $v_{impact}$ (and thus $\eta$) distribution is unknown. We use our simulation results to obtain a normalized distribution of $v_{impact} / (q \times v_{escape})$ and fit it to a continuous probability density function (PDF; $f(\eta, q)$). Scaling by $v_{escape}$ allows us to directly use the mass scaling laws and scaling by $q$ helps compare simulation data between different excitations. To reduce the effect of outliers, we combine the $v_{impact} / (q \times v_{escape})$ data from all simulations before fitting. Because $q$ changes with mass of the satellite, we normalize a given $v_{impact}$ by the $q$ value at that time of the simulation. We can now average the mass loss $\delta M(\eta)$ over $\eta$ to remove the dependence on $\eta$:

\begin{equation} \label{eqn:averaged deltaM}
        \langle \delta M(q) \rangle =\int_{0}^{\infty} \delta M(\eta) f(\eta, q) d\eta
\end{equation}
Explained in more detail in section \ref{sec:breakup timescale results} and seen in \cref{fig:pdf fit to v_impact}, the best fit PDF is a log-normal distribution which has the form:

\begin{equation} \label{eqn:lognormal form}
    f\left(\eta, q\right) = f\left(\eta, s, \sigma (q)\right) = \frac{1}{s \eta \sqrt{2 \pi}}\exp\left[{-\frac{\ln^2\left(\eta/\sigma\right)}{2s^2}}\right]
\end{equation}
for the value $\eta$ where $s$ is the shape parameter and $\sigma$ is the scale parameter that accounts for $q$ in each simulation, i.e. $\sigma = q \times \sigma_{fit}$. We fit $s$ and $\sigma_{fit}$ to simulation data. 

Combining all this together, we calculate a mass loss rate. The mass loss is inversely proportional to re-impact timescale $\tau$ (\cref{eqn:tau simplified}), and directly proportional to the mass lost per impact (\cref{eqn:scaling laws form}). For the latter, we scale the target body's mass loss per impact by that impactor's mass ($m_i$). Summing over each impact $i$ we get:
\begin{align} \label{eqn:timescale derivation step 1}
    \frac{dM}{dt} &\propto - \frac{\sum_{i} \left( m_i \times \delta M(\eta_i) \right)}{\tau}
\end{align}
The negative sign signifies mass loss. To simplify calculations, we can average out the mass loss over $\eta$ (\cref{eqn:averaged deltaM}) and use the total impactor mass $M_i$. Because $\eta$ is a function of $q$ (\cref{eqn:eta sim q}), we have $dM/dt$ as a function of $q$.
\begin{align}
\frac{dM}{dt}(q) &\propto - \frac{\langle \delta M(q) \rangle \sum_{i} m_i}{\tau} \label{eqn:timescale derivation step 2} \\
\frac{dM}{dt}(q) &\propto - \frac{\langle \delta M(q) \rangle}{\tau} M_i \label{eqn:timescale derivation step 3}
\end{align}
Because the impactors are coming from the target moon, any mass lost from the moon is fed into the impactors creating a feedback loop. Therefore, we can define the total impactor mass $M_i(t) = M_{i, t = 0} + M_{t = 0} - M(t)$ where $M(t)$ is the target moon's mass. Substituting in $M_i(t)$ and \cref{eqn:tau simplified}  gives us differential equation for $M(t)$ that we can now solve.

\begin{align} 
\frac{dM}{dt}(q) &\propto - \frac{\langle \delta M(\eta) \rangle}{K} (M_{i, t = 0} + M_{t=0} - M) M \label{eqn:timescale derivation final}
\end{align}


\section{Results} \label{sec:results}

\begin{figure}
    \centering
    \plotone{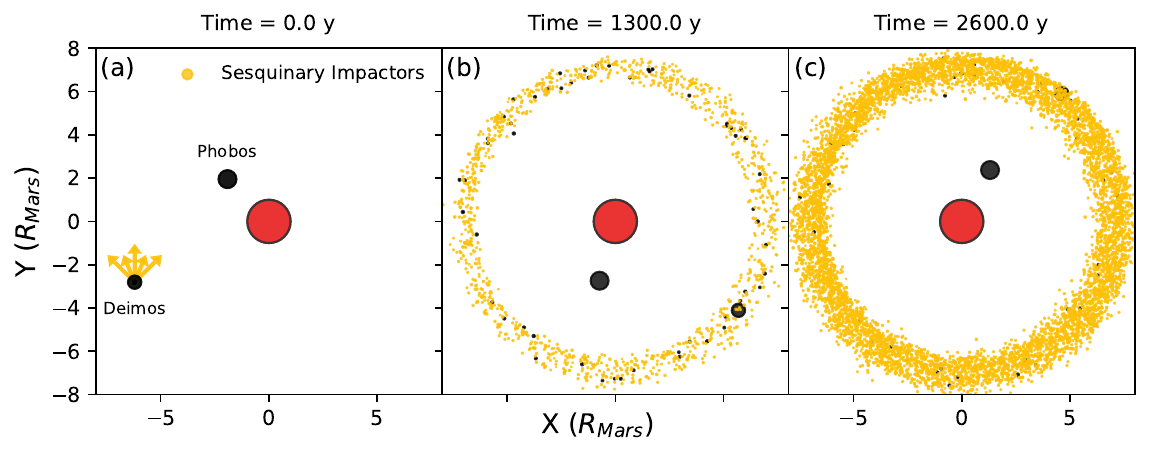}
    \caption{Typical sesquinary catastrophe behavior in a Mars-centric frame. Panel (a) shows the initial simulation setup described in section \ref{sec:sim set up}. Deimos (and therefore the ejecta) is given an excitation $q = 0 - 40$ by giving it a higher $e$ and/or $i$. Ejecta from Deimos has an initial velocity $v = 1-3 \ v_{escape}$ with total ejecta mass $M_{impactor} = 10^{-3} \ M_{Deimos}$ unless mentioned otherwise. The gravitational harmonics for Mars from GMM3~\citep{Genova2016SeasonalScience} are also included up to degree and order 6. Repeated sesquinary impacts quickly fill up the region of Deimos and create an ejecta debris ring. The high-velocity re-impacting ejecta contribute to this runaway sesquinary erosion. Phobos, Deimos, and impactor particles are scaled up for visual ease.}
    \label{fig:sesquinary sim plots}
\end{figure}

Evolving the system as set up in panel (a) of \Cref{fig:sesquinary sim plots} shows that the sesquinary catastrophe erodes an excited Deimos. \Cref{fig:sesquinary sim plots} shows the typical behavior of the mechanism with sesquinary impactors progressively filling up the region of Deimos and shaving mass off it. The orbital elements of Deimos show some stochastic evolution with no clear behavioral trend across the various simulations, preventing us from drawing clear conclusions about the orbital evolution. For almost all runs, the sesquinary excitation $q$ of Deimos stays about the same with a little increase in eccentricity due to the excited impacts. Regardless, orbital element changes are outpaced by the mass change from sesquinary break-up.

\subsection{The sesquinary catastrophe can completely erode Deimos} \label{sec:sesquinary results}

\begin{figure}
    \centering
    \plotone{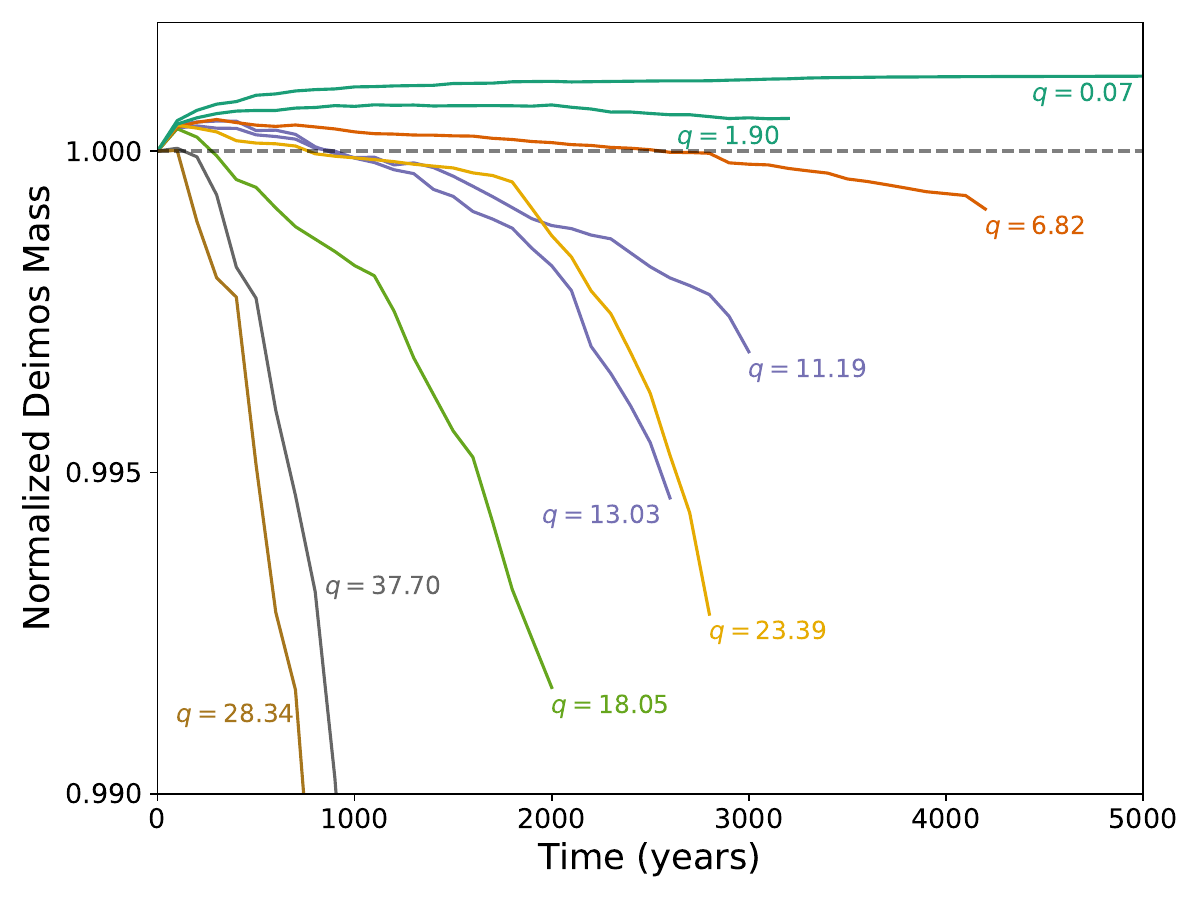}
    \caption{Mass evolution of Deimos in the sesquinary catastrophe vs time for various initial excitations via N-body simulations. This shows large-scale mass loss due to the sesquinary catastrophe across a range of excitations $q$, proportional to the excitation level. In the early stages, we see some accretion because not all particles have precessed enough out of alignment to gain erosive impact velocities. We show selected simulations for visual ease, but about $90$ simulations were run that show similar behavior.}
    \label{fig:mass loss sims vs q}
\end{figure}

To test if breakup is possible by the sesquinary catastrophe, we started Deimos at different excitations with a higher initial $e$ and/or $i$, effectively testing $q = 0 - 40$. \Cref{fig:mass loss sims vs q} shows mass loss at various excitations of $q \gtrsim 5$ due to the sesquinary catastrophe on Deimos. Visually, the mass loss also scales with $q$ albeit with some stochasticity in the simulations. The main sources of the stochasticity are the monte carlo aspect of collisional outcomes in collisional fragmentation model FRAGGLE in Swiftest~\citep{Wishard2023Swiftest:Systems} and the differences in impactor properties (such as velocity and mass) due to differential precession. For example for the latter, a given simulation may have larger and faster impactors hit Deimos early on causing greater mass loss, while others may have smaller impacts that cause a more gradual initial mass loss. These changes look more prominent because we are only able to model a small part of the erosion. We can still conclude that the sesquinary catastrophe can erode a small close-in planetary moon at various values of excitation parameter $q$.

Because of computational constraints, we also ran accelerated simulations where Deimos was started at lower initial masses to mimic the later stages of sesquinary breakup with the remaining mass put into a limited number of larger impactors. The set up in the same as in section \ref{sec:sim set up} except that we start with $300$ particles in the debris disk with the total mass ``removed" from Deimos spread among the debris. Because the total debris mass varies depending on the initial mass of Deimos, the individual particle radii vary from $\sim 345$~m to $\sim 664$~m for the $M_{initial} = 0.95 \ M_{Deimos}$ to $0.7 \ M_{Deimos}$ respectively. The underlying behavior is still the same but each impact is more efficient at causing mass loss because of the bigger size (see section \ref{sec:semi-analytical timescale}). These piecewise simulations mimic impactor replacement and helps us accelerate sesquinary breakup behavior to ensure that complete breakup is seen. 

\begin{figure}
    \centering
    \plotone{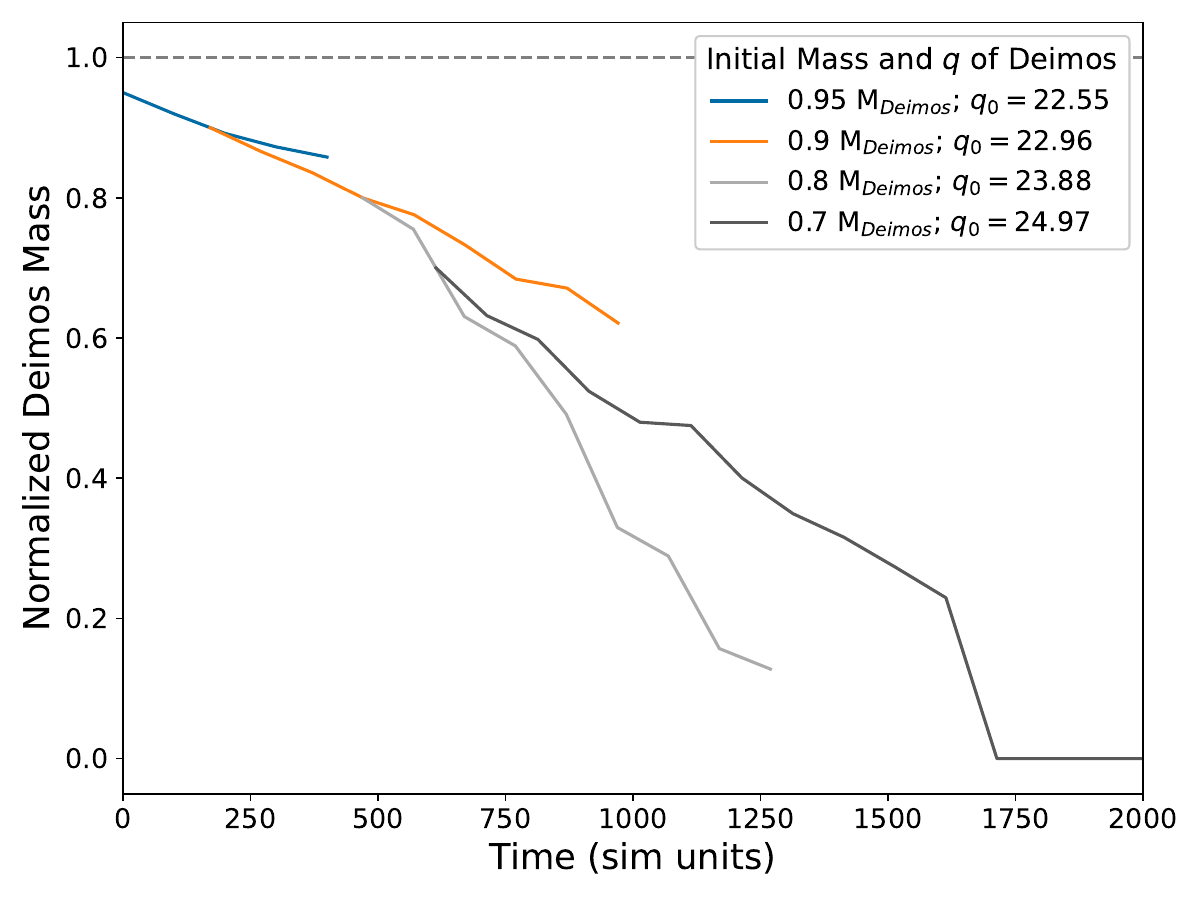}
    \caption{Mass evolution of Deimos at initial $e = 0.05$ and $i = 5^\circ$ in the sesquinary catastrophe vs time at different initial mass steps via accelerated N-body simulations. The mass lost from Deimos is put into large impactors so that total mass of the system is the same. This accelerates the simulations with more mass loss per impact, and qualitatively show that complete breakup is possible. NOTE: ``sim units" in each simulation are based on years. We present the time in ``sim units" rather than years to emphasize that each simulation has a different time scaling because of the inflated impact rate, and the remnant mass is not a continuous function of time.}
    \label{fig:mass loss sims with mass steps}
\end{figure}

For these simulations, Deimos was given an $e = 0.05$ and $i = 5^\circ$ ($q \simeq 22$ at modern-day mass) to test sesquinary breakup at an excitation level far enough away from the theoretical tentative breakup threshold of $q \sim 5$ from \citet{Hyodo2020EscapeEjecta}. Because of the slow orbital element change of Deimos, these stepped simulation can be considered directly equivalent to each other to a large degree. The results shown in \Cref{fig:mass loss sims with mass steps} show that Deimos can break up completely due to the sesquinary catastrophe and has a very fast demise once Deimos reaches $99 - 95\%$ of its present-day mass.

\subsection{Break-up Timescale} \label{sec:breakup timescale results}

\begin{figure}
    \centering
    \plotone{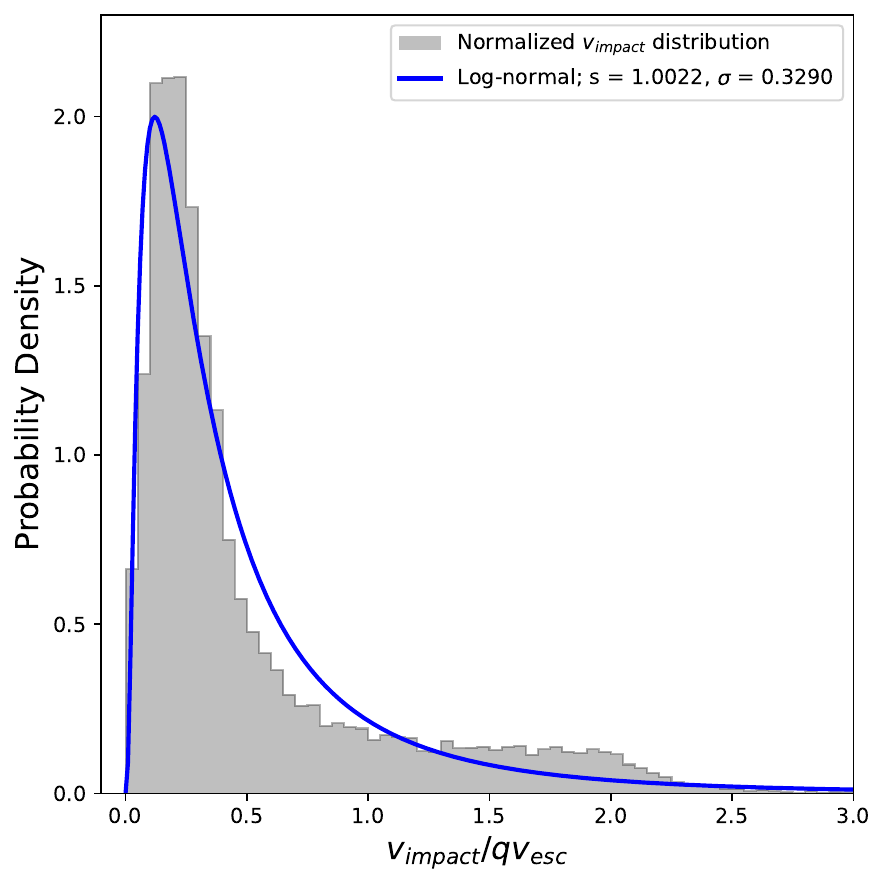}
    \caption{Probability density function (PDF) fits to the normalized $v_{impact}$ data. We combine and normalize the impactor velocity ($v_{impact}$) data across all simulations and fit the distribution to multiple PDFs using SciPy~\citep{Virtanen2020SciPyPython}. Here is the best fit log-normal distribution with the fitting parameters in the legend.}
    \label{fig:pdf fit to v_impact}
\end{figure}

To estimate a break-up timescale from the sesquinary catastrophe, we normalize and fit the combined $v_{impact}$ distribution from our simulations to various known continuous distributions. We tested fits to a log-normal, maxwell, rayleigh, gamma, beta, and more distributions. For simplicity, we chose to fit just one PDF as a first-order estimate. While a combination of multiple PDFs may fit better and be more representative of the data, it is a more complicated fit and is left for future work. As shown in \Cref{fig:pdf fit to v_impact}, a heavy tailed distribution like log-normal with shape parameter $s = 1.002152$ and scale $\sigma_{fit} = 0.328925$ fits the best to the combined $v_{impact}$ data. We plug these fit parameters into the log-normal distribution (\cref{eqn:lognormal form}) which in turn is used to average the mass loss per impact (\cref{eqn:averaged deltaM}). We then carry out the procedure described in section~\ref{sec:semi-analytical timescale} and numerically integrate \cref{eqn:timescale derivation final} in Mathematica~\citep{WolframResearch2024Mathematica} to obtain a breakup timescale for Deimos depending on initial $q$. 

Because of the decaying exponential behavior of mass, we calculate the time for breakup to $0.01 \ M_{Deimos} = 1\%$ of Deimos' original mass when starting with total impactor masses of $10^{-4}$ and $10^{-3} \ M_{Deimos}$. The total initial impactor mass of $10^{-4} \ M_{Deimos}$ is the expected amount of mass expelled from Deimos from a Voltaire-like impact~\citep{Nayak2016EffectsGeology} while $10^{-3} \ M_{Deimos}$ is the typical starting debris mass in our simulations. 
We have to choose an end point for Deimos' mass for a couple of reasons. Constraining the mass threshold of when a collisional remnant can still be classified as the original body is tricky. We saw a complete break up of the original ``Deimos" via a supercatastrophic disruption~\citep[target body loses $> 50\%$ of its mass,][]{Leinhardt2012COLLISIONSLAWS} at around $\sim 0.25 - 0.15 \ M_{Deimos}$ in our simulations as the bodies approach the similar-mass collision regime. In addition to this approximate range, the breakup timescale has a decaying asymptotic behavior. We pick 2 values where the final mass of Deimos is $0.01 M_{Deimos}$ (or $1\%$ of its original mass) and $0.1 M_{Deimos}$ (or $10 \%$ of its original mass). The amount of time it takes to get to $0.1 M_{Deimos}$ is only $\sim 4 \%$ lesser than the amount of time it takes to get to $0.01 M_{Deimos}$ for a given $q$ value. For that large of a mass change, the time difference is very small and the final time for breakup does not change much. Therefore, the final mass value of $1 \% = 0.01 \ M_{Deimos}$ is lower than what might be considered a complete break up but provides an upper bound on the breakup time. 

\begin{figure}
    \centering
    \plotone{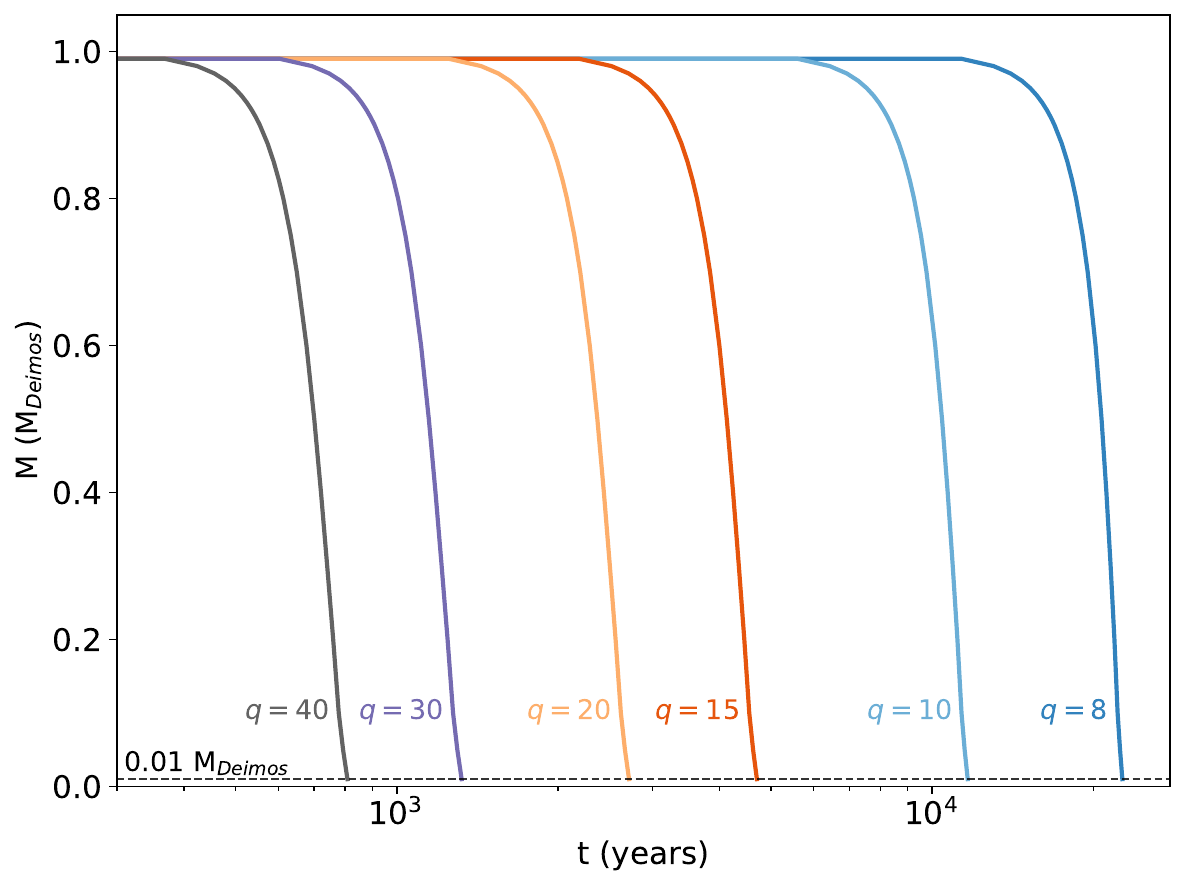}
    \caption{Semi-analytical mass vs time for Deimos as a function of initial $q$. Given an initial excitation $q$, we can calculate the mass evolution over time. There is a slow erosion of the excited Deimos until it reaches the ``tipping point" of $0.99 - 0.95 \ M_{Deimos}$, after which the erosion is very quick because of the increased flux of impactors. Of the total time for break-up, $\sim 50 \%$ of it is spent reaching $0.99 \ M_{Deimos}$ and then next $\sim 50 \%$ is spent reaching the final value of $0.01 \ M_{Deimos}$. }
    \label{fig:semi-analytical mass vs time}
\end{figure}

\begin{figure}
    \centering
    \plotone{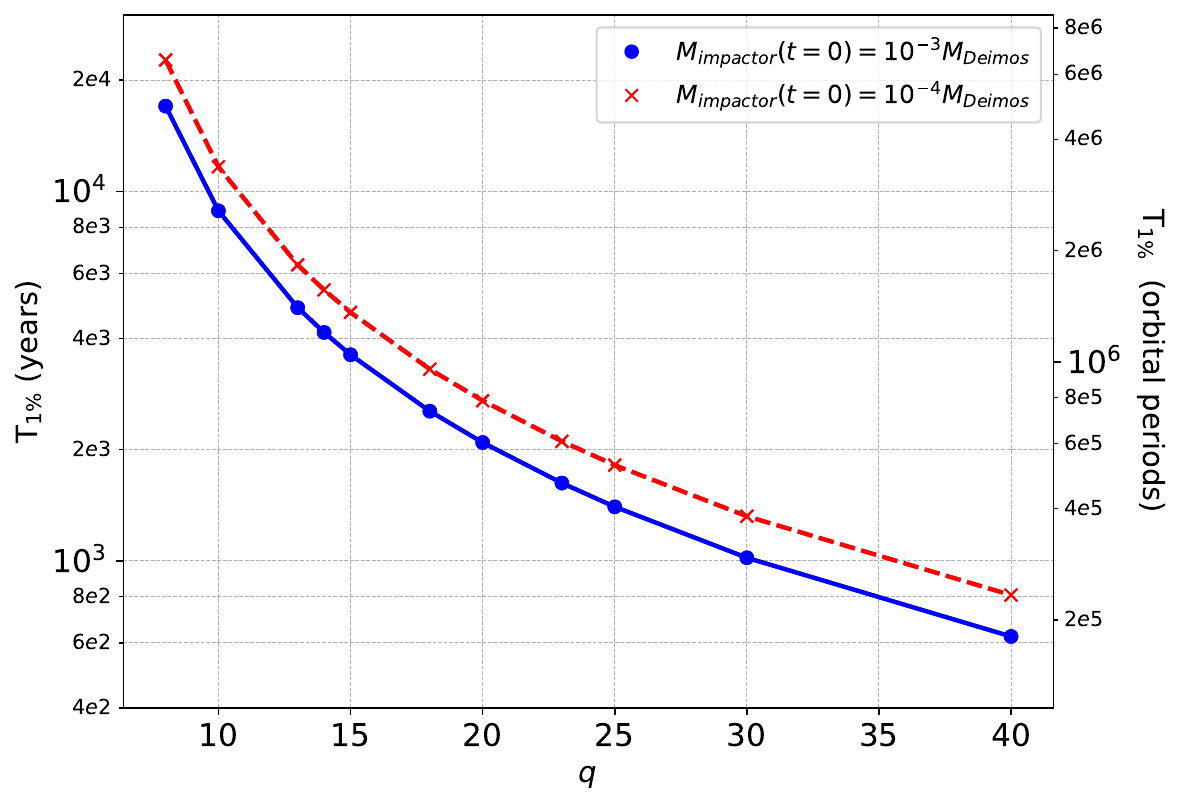}
    \caption{Semi-analytical break-up time for Deimos as a function of initial $q$. Here, we calculate the time taken by the sesquinary catastrophe to erode Deimos to $0.01 \ M_{Deimos}$ or $1 \%$ of its original mass for a given initial excitation value $q$. We use the process described in section \ref{sec:semi-analytical timescale} to get an averaged estimate of break-up time. The break up time is on the order of $\sim 10^3$~years $ \simeq 10^6$~orbits for Deimos. The initial impactor mass $M_{impactor}(t=0)$ has some effect effect on break-up time, but keeps the same order of magnitude. An initial $M_{impactor} = 10^{-3} \ M_{Deimos}$ takes $\sim 22 \%$ lesser time for break-up than $M_{impactor} = 10^{-4} \ M_{Deimos}$.}
    \label{fig:breakup timescale results}
\end{figure}

The semi-analytical evolution of Deimos' mass over time under the sesquinary catastrophe is shown in \Cref{fig:semi-analytical mass vs time}. The breakup time is on the order of few $10^3$~years or $10^6$~orbits for Deimos (\Cref{fig:breakup timescale results}). We see an extremely slow erosion of the moon until $0.99 \ M_{Deimos}$ (about half the total time) and then a very sharp drop off until complete breakup. This sharp drop in mass is aided by the increased amount of available impactors and the increased excitation $q$ of the moon as it is eroded.

\subsection{Sesquinary excitation threshold for breakup} \label{sec:breakup threshold}

Theoretically, the scaling laws from~\citet{Hyodo2020EscapeEjecta} predict an impact angle-averaged $v_{impact} \gtrsim 5 \ v_{escape}$ for the erosive impacts on a strengthless body. This corresponds to a sesquinary excitation of $q \gtrsim 5$ for erosive impacts, as corroborated by the expected approximate breakup threshold of $q \sim 10$ in \citet{Cuk2023SesquinaryOrbits} and our simulations. However, since these are approximate ranges, we can use the semi-analytical approach in sections \ref{sec:semi-analytical timescale} and \ref{sec:breakup timescale results} to account for the impactor velocity spread in simulations. We obtain a breakup threshold of $q = 5.66$ from semi-analytical calculations. This would imply that modern-day Deimos~\citep[$q = 7.6$,][]{Cuk2023SesquinaryOrbits} is undergoing and susceptible to a sesquinary catastrophe.

However, modern-day Deimos does not have an observed dust ring at current observational limits~\citep{Showalter2006ATelescope, Krivov2006SearchOpportunity, Zakharov2014DustSpace} and is not being broken apart. However, potential observations of the expected dust ring or torus~\citep{Liu2020ConfigurationTrajectories, Krivov1997MartianDiscovery, Krivov2006SearchOpportunity, Soter1971TheMars, Hamilton1996TheMars} by MMX will improve our understanding~\citep{Kobayashi2024AdvancementsMMX., Kobayashi2018InSensor}. The lack of dust ring implies that Deimos is below (though it may be close to) the sesquinary catastrophe threshold at its current configuration.

The strength of the body, radiation forces acting on small debris, and other effects would likely alter this threshold. A body with more strength would be more resistant to breakup, take longer, and move the plot in \cref{fig:breakup timescale results} to the right. Our cratering collisional model based on \citet{Hyodo2020EscapeEjecta} neglects strength, and thus sets a lower bound on the breakup threshold $q$ for strength-less bodies. Radiation forces drive small debris away, reduce the amount of impactors available, and thus also increase the threshold by requiring more excited impactors to facilitate similar breakup behavior (see section \ref{sec:caveats} for discussion). Therefore, material strength and radiation are inclusions that need to be studied better and factored into this threshold when characterizing it. In our calculations, we see extremely sharp asymptotic behavior below $q = 8$ and the mass loss rate slows down quickly. The expected time for breakup for Deimos starting at $q = 6$ is $6\times$ longer than that for $q = 8$. Transient artifacts from the numerical integration and interpolation have a much larger effect with these small number mass loss rates. Combining all of these factors above, we consider a threshold of $q \simeq 8$ as a safe value for the sesquinary catastrophe threshold.

\section{Discussion} \label{sec:discussion}


The sesquinary catastrophe is the first step in solving the problem of a past excited orbit of Deimos by breaking the excited moon apart. This is a model-independent process that only requires an excited moon with Deimos-like mass and semi-major axis. If Deimos was excited in the past as expected (section \ref{sec:background}), this is a potential avenue to remove its dynamical excitation. We do not expect Deimos to be excited by Phobos in the future due to a lack of strong mean motion resonances between the two as Phobos evolves tidally inwards. For reference, the 3:1 MMR is at $3.33$~$R_{Mars}$ while Phobos is currently at $a = 2.762$~$R_{Mars}$~(\cref{tab:moons orbital el}).

The sesquinary catastrophe is akin to sandpapering. Sesquinary impactors slowly shave small amounts of mass, on the order of few impactor masses, off of Deimos per impact. As a consequence of this process, smaller particles are created with each impactor-Deimos and impactor-impactor collisions eventually leading to small debris particles in the final debris disk. We see similar particle size behavior in our simulations as well, though the exact grain-size limit is not modeled in this work. Collisional fragments in these simulations are only generated down to the user-set minimum fragment size of $\sim 100$~m. We expect easier circularization of smaller debris particles via collisional damping. 

Once the moon is broken up, we expect that all information about the original body is lost. The broken-up moon will be debris in a larger ring of debris with the same origin. This makes identifying the original body difficult. There might be some larger pieces or chunks, but the main idea stays the same. As the debris orbit and collisionally lose excitation, the material will be more mixed up because of differential precession. This results in a dynamically cool debris ring with well mixed material. For the debris to re-accrete into a single large body, the debris ring has to be de-excited below the sesquinary breakup threshold, otherwise any formed large body will be eroded again by the excited debris. Therefore, regardless of initial excitation, an excited debris disk has to dynamically cool below the sesquinary breakup threshold to be able to re-accrete. As the re-accretion starts below the same excitation limit, constraining the past of the re-accreted body is extremely difficult. This expected re-accretion into a less excited body would complete the process of removing the dynamical excitation of past Deimos.

However, our models are not the appropriate tools for constraining a limit on the particle size, nor the next step of re-accretion, and that is left for future work.

\subsection{Observational signatures} \label{sec:observational signatures}

With the impending Martian Moon eXploration mission (MMX)~\citep{Kuramoto2022MartianPlanets}, observational signatures that test this theory are needed. The following signatures point to a possibility that Deimos went through the sesquinary catastrophe in the past, rather than definitively prove that it did. What the sesquinary catastrophe hypothesis definitively suggests is that any potential excitation past about $q \sim 8$ would be erased by the catastrophe, so the present-day orbit is not a strong constraint on (potential) past excitation events.

The first signature would be that Deimos is a rubble-pile moon with relatively small grains. Because of the expected progressively smaller size of ejected particles, the resultant debris disk would be composed of finer grains\footnote{We do not aim to set a specific grain-size constraint as we do not accurately model particle sizes. We only speculate a smaller than typical (ex: big-boulders) material make-up.} that re-accrete into modern-day Deimos. This would be consistent with the smooth surface of Deimos seen in images from the Viking missions~\citep{Veverka1977VikingResults, Duxbury1978SpacecraftDeimos}\footnote{https://science.nasa.gov/mission/viking-2/ and https://nssdc.gsfc.nasa.gov/imgcat/html/object\_page/vo2\_423b63.html; Accessed on 7/1/2025}, its somewhat ellipsoidal shape, low mass, and porous structure. 

Secondly, Deimos should be largely homogeneous. The debris disk material should be thoroughly mixed after many precession cycles and then re-accrete into a homogeneous body. Recent moment of inertia fits to Phobos and Deimos are consistent with bodies of uniform density~\citep{Brozovic2025RevisedDeimos}. This is consistent with our expected results. The debris disk material would be exposed to cosmic rays and solar radiation before re-accretion. This will be difficult to test because of unconstrained exposure times of the debris disk and current space weathering of Deimos' surface. A sample return mission would be able to test the geologic make-up of Deimos the best.

\subsection{Caveats} \label{sec:caveats}

Some caveats of our work are that we do not include non-gravitational effects such as radiation. Radiation effects on small particles around Mars are significant with eccentricity and inclination variability, and lifetime implications and will affect the debris disk after break-up~\citep{Burns1979RadiationSystem, Hamilton1996TheMars, Hamilton1996CircumplanetaryElectromagnetism, Liu2020ConfigurationTrajectories, Liang2023Giga-yearMars}. The lifetime of dust particles near Deimos depends on the size but larger particles ($R \gtrsim 10 \ \mu m$) last in the system for $\sim 10^4$~years~\citep{Liu2020ConfigurationTrajectories} and up to $10^9$~years~\citep{Liang2023Giga-yearMars}. Dust particles near modern-day Deimos see eccentricity and inclination increases up to $0.3$ and $12^\circ$ respectively~\citep{Hamilton1996TheMars} that may aid the erosion process. At higher $q$ values, more mass is lost per impact and the re-impact time $\tau$ is also lower. The expected lifetime of debris particles near Deimos \citep[$10^{4-9}$~y,][]{Liu2020ConfigurationTrajectories, Liang2023Giga-yearMars} is longer than the expected re-impact times for excited sesquinary impactors \citep[$\tau = 1600$~y at $q = 7.6$,][]{Cuk2023SesquinaryOrbits} and breakup time for Deimos from the sesquinary impacts ($10^{3-4}$~y). As a result, while there is a competition between radiation loss and sesquinary impacts for small dust grains, non-gravitational effects do not seem to affect the large scale process at hand. Depending on the size frequency distribution of the debris, non-gravitational forces may alter the sesquinary catastrophe threshold, leading us to choose a value of $q \simeq 8$.

Tidal effects were not included in this work for a couple of reasons. Our simulations are run on $10^3$~year timescales, and this is very short when accounting for tidal effects which have $10^6$~year timescales~\citep{Yoder1982TidalPhobos, Black2015TheSystem, Cuk2025TwoPhobos}. While Deimos is outside the synchronous orbit and is expected to move away from Mars, it is very close to it, and thus sees effectively no change in orbit over our timescales. Secondly, Phobos is tidally evolving towards Mars much faster than Deimos, but still evolves on the order of $10^6$~years~\citep{Yoder1982TidalPhobos, Black2015TheSystem, Cuk2025TwoPhobos}. We do include gravitational harmonics that both affect the orbits on short timescales and contribute to differential orbital precession between the bodies.

We comment earlier that the progressive collisions create smaller fragments. We see this behavior in our simulations but cannot constrain a particular size frequency distribution. The collision models used in Swiftest~\citep{Wishard2023Swiftest:Systems, Leinhardt2012COLLISIONSLAWS, Hyodo2020EscapeEjecta} are constrained for large planetary bodies in the gravity regime ($R \gtrsim 100$~m at $\rho = 1 \ g/cm^3$,~\citet{Stewart2009Velocity-DependentPlanetesimals}). As a result, the models used here can be used to show general behavior, but are not the appropriate tools for constraining the size of small collisional fragments. Results from MMX~\citep{Kuramoto2022MartianPlanets} with more precise modeling will help constrain the past of Deimos, including the likelihood of the sesquinary catastrophe, better.

\subsection{Adrastea and Thebe as potential laboratories of the sesquinary catastrophe} \label{sec:adrastea and thebe}

The sesquinary catastrophe can be applied to other moons in the solar system as well, as discussed in \citet{Cuk2023SesquinaryOrbits}. An example of this process occurring could be Jupiter's moon Adrastea with a $q \sim 8.5$ and $\tau \sim 910$~years and/or Thebe with a $q \sim 18.0$ and $\tau \sim 8900$~years~\citep{Cuk2023SesquinaryOrbits}. This high $q$, low $\tau$, and lack of a resonance at Adrastea and Thebe are ideal materials for the sesquinary catastrophe. Material is currently coming off of Adrastea and Thebe at a variety of particle sizes that feeds Jupiter's ring~\citep{Burns2004JupitersSystem, Ockert-Bell1999TheExperiment, Burns1999TheRings}. Thebe's Gossamer ring is the least dense ring and implies that it is not losing material at a faster rate than the others~\citep{Cuk2023SesquinaryOrbits, Ockert-Bell1999TheExperiment, Hamilton2008TheShadow}. This means that Thebe may have significant strength or be in the very early stages of a sesquinary catastrophe. Adrastea also likely has material strength~\citep{Tiscareno2013CompositionsDensity}. While this material is thought to be from heliocentric impactors due to Jupiter's large gravity well~\citep{Burns1999TheRings}, both moons are very small and provide a small collisional cross-section. A short-timescale process, like the sesquinary catastrophe, is likely more feasible at explaining the continual ejection of material. Radiation and Lorentz forces help propagate the material away as well, but the potential of the sesquinary catastrophe means that the system does not need continuous heliocentric impacts and is self-propagating once started. Regardless, this is a very complicated and degenerate system. The mass shedding from Adrastea and Thebe is likely a combination of sesquinary impactors, heliocentric impacts, tidal breakup, non-gravitational forces, and the hill sphere being close to the surface~\citep{Burns2004JupitersSystem}. 

\section{Conclusions} \label{sec:conclusions}

In this work, we numerically test the concept of sesquinary catastrophe on Deimos and show a working proof that this is a viable mechanism to break apart a dynamically excited planetary moon, paving the way for its re-accretion on a dynamically cold orbit. This mechanism is a model-independent way to potentially reconcile the hypothesized excited past of Deimos (section \ref{sec:background}) with its dynamically cooler present.

\begin{itemize}
    \item A runaway cascade of sesquinary impactors (initially initiated by a heliocentric impact) can grind down and break up a dynamically excited, small, close-in planetary moon. If Deimos was excited in the past beyond $q \sim 8$, it would have undergone the sesquinary catastrophe, been broken up into a debris disk, and re-accreted on a dynamically cooler orbit. The breakup time would be on the order of few $10^3$~years or $10^6$~orbits.
    \item Because this process is a slow grinding down of the moon, akin to sandpapering, subsequent collisions create smaller and smaller particles. We expect Deimos to be a re-accreted sand-pile moon. This would explain its smooth surface and atypical shape with a homogeneous mixing of source material.
    \item Almost all information on the prior orbital state is lost after sesquinary breakup, with the sole exception of orbital angular momentum. While the material of the previous broken-up moon is still conserved, the exposure of small particles to radiation may alter the material. This makes constraining the past of Deimos, or any such moon, extremely difficult.
    \item While most applicable to an ancient Deimos, this could apply to other moons that are gravitational aggregates with Deimos-like strength with an orbital excitation $q \gtrsim 8$ such as Jupiter's moons Adrastea and Thebe. This threshold was obtained from numerical integration of averaged impact velocity statistics and does not include non-gravitational effects that may alter the threshold for sesquinary catastrophe.
\end{itemize}

\begin{acknowledgments} \label{sec:acknowledgments}

This work was funded by NASA Emerging Worlds Program award 80NSSC23K1266. We would like to thank Govardan Gopakumar for helping with fitting the impact velocity data. We also thank two anonymous reviewers for greatly improving the manuscript.

\end{acknowledgments}

\begin{contribution}

KA was responsible for writing, editing, and submitting the manuscript, developed the cratering collision model, ran appropriate simulations, compiled, and analyzed results. MC came up with the initial research concept, is the PI of the grant, edited the manuscript, supervised the analysis, and advised KA. DM is the co-I on the grant, maintains the Swiftest github, supervised code development by KA, edited the manuscript, and is the primary advisor for KA.


\end{contribution}

%

\software{Swiftest~\citep{Wishard2023Swiftest:Systems}, Mathematica~\citep{WolframResearch2024Mathematica}, SciPy~\citep{Virtanen2020SciPyPython}, NumPy~\citep{Harris2020ArrayNumPy}}
\facilities{Purdue Rosen Center for Advanced Computing Clusters: Negishi and Bell}


\appendix

\section{Appendix information}


\bibliography{references}{}
\bibliographystyle{aasjournalv7}



\end{document}